\documentclass[aps,prb,twocolumn,10pt,superscriptaddress,showpacs,amsmath,amssymb]{revtex4-1}

\usepackage{multibib}

\usepackage{graphicx}
\usepackage{latexsym}
\usepackage{amsmath}
\usepackage{amssymb}
\usepackage{amsfonts}
\usepackage{color}
\usepackage{bm}
\usepackage{verbatim}
\usepackage{color}

\begin{document}

\title{Comment on ``Ginzburg-Landau theory of two-band superconductors: Absence of type-1.5 superconductivity" by
V. G. Kogan and J. Schmalian }

\author{Egor Babaev}
\affiliation{Physics Department, University of Massachusetts, Amherst,
 Massachusetts 01003, USA}
\affiliation{Department of Theoretical Physics, The Royal Institute of Technology, 10691 Stockholm, Sweden}
\author{Mihail Silaev}
\affiliation{Department of Theoretical Physics, The Royal Institute of Technology, 10691 Stockholm, Sweden}
\affiliation{Institute for Physics of Microstructures RAS, 603950
Nizhny Novgorod, Russia.}

\begin{abstract}
The recent paper by V. G. Kogan and J. Schmalian Phys. Rev. B {\bf
83}, 054515 (2011) argues that the widely used two-component
Ginzburg-Landau (GL)  models are not correct, and further
concludes that in the regime which is described by a GL theory
there could be  no disparity in the coherence lengths of two
superconducting components. This would in particular imply that (in
contrast to $U(1)\times U(1)$ superconductors), there could be no
``type-1.5" superconducting regime in $U(1)$ multiband
systems for any finite interband coupling strength. We
point out that these claims are incorrect and based on an
erroneous scheme of reduction of a two-component GL theory. {\it Note added}: below we also attach a separate rejoinder on reply by Kogan and Schmalian.
\end{abstract}

 \maketitle

\section{Introduction}

The recent works Refs. \onlinecite{kogan, kogan1} claim that the
two-component Ginzburg-Landau (TCGL) theories can not be used to
address any properties of two-component superconductors which
involve disparity of density variations, in particular to describe
type-1.5 superconducting state where $\xi_1
<\sqrt{2}\lambda<\xi_2$
\cite{bs1,GL2mass,GL2mass2,1type-15,2type-15,semimeissner,Silaev,Silaev2,recent1,recent2,recent3,recent3b,daonew,milosevic2,3bands,review}.
Here we point out several crucial errors in the analysis
 \cite{kogan,kogan1} thus demonstrating  that the following main points in
Ref.\onlinecite{kogan} are incorrect,
\begin{itemize}
\item
{ attempts to employ the GL
functionals, on the one hand, and to assume different length
scales, on the other, cannot be justified.}
\item
{ the idea
of 1.5-type superconductivity is not warranted by the GL
theory}
\item
{ $\Delta_1({\bf r},T)/\Delta_2({\bf r},T)=const$,
... this ratio remains the same at any T in
the GL domain}
\end{itemize}

First let us note that   Refs. \onlinecite{kogan, kogan1} fail to
distinguish between two  classes of systems where the type-1.5
state was previously discussed (i) $U(1)\times U(1)$   and (ii) two-band
superconductors where interband coupling explicitly breaks
symmetry down to $U(1)$. Namely  the Refs. \onlinecite{kogan,
kogan1} mix up various aspects of physics specific to $U(1)\times
U(1)$  from Ref. \onlinecite{bs1} (such as the very definition of the coherence
lengths) with the different in several respects physics of $U(1)$
systems. The definitions of coherence lengths and type-1.5 regime
in systems with non-zero interband coupling were explicitly
discussed  in detail in Ref. \onlinecite{GL2mass}, long before the
appearance of Refs. \onlinecite{kogan,kogan1}. Thus  claims in
Refs. \onlinecite{kogan,kogan1} that this coupling was neglected in works on type-1.5 superconductivity
are factually incorrect. Note that  $U(1)\times U(1)$ symmetry is also possible in
superconductors \cite{LMH,ns} which represents the most
straightforward example of systems which cannot be characterized
by a single universal GL parameter (physical examples can be found
in Refs.\onlinecite{bs1,GL2mass2}). However in what follows we
focus exclusively on $U(1)$ two-band superconductors.

Two-component GL (TCGL) model were derived microscopically in
Refs. \onlinecite{Tilley,gurevich,zhitomirsky}. However indeed the
conditions under which two-component  GL expansions for two-band
superconductors are formally justified, were not known (to the
best of our knowledge) at the time of publication of
Refs.\onlinecite{kogan,kogan1} but
 were  rigorously established
recently \cite{Silaev2}. Therefore the aforementioned claim that
TCGL expansion is injustifiable \cite{kogan,kogan1} is incorrect.
In this comment we  discuss which incorrect assumptions and
technical errors  led the authors of
Refs.\onlinecite{kogan,kogan1} to opposite conclusions.

\section{ Definitions of the GL regime in application to $U(1)$ multiband systems.}

Let us start with definitions. The Refs. \onlinecite{kogan,
kogan1} defines ``GL theory"   as the free energy proportional to
$\tau^2$  and the modules of the fields varying as
$\tau^{1/2}$ where $\tau=(1-T/T_c)$. Such
simplistic definition indeed can be encountered
in books on superconductivity which consider simplest
single-component systems. However unfortunately
such a definition does not work in
general.
The Ginzburg-Landau theory is a more general concept
of a classical field theory description
of a system, which in many physical cases does not necessarily appears in the leading order
$\tau$-expansion.
In particular such a definition
 contradicts all
existing literature on multicomponent GL theories in two band
superconductors, which in contrast adopts the more general
definition of GL expansion of the free energy by powers of gap
amplitudes and spatial gradients
\cite{Tilley,gurevich,zhitomirsky,multicomp}.
It should be noted that it most obviously follows from the $U(1)$
symmetry of two-band superconductors, that the leading order
expansion in the parameter $\tau$ yields a {\it single} order
parameter field characterized by a single coherence length {\it by
construction} (see e.g. a standard textbook Ref.\onlinecite{pathria}). The works
\cite{Tilley,gurevich,zhitomirsky}, as well as more recent paper
\cite{Silaev2} use more general expansion in powers of gradients
and amplitudes of the two gap functions $\Delta_{1,2}$, which most obviously yields a
more complicated temperature dependence, and  cannot be expected
to be obtainable in leading order expansion in $\tau$.
 { The TCGL expansion  for two-band
superconductor is thus an example of an expansion
in several small parameters (small gaps and gradients).
However this fact does not make it unjustifiable as claimed \cite{kogan}.
Indeed recently it was justified on formal grounds
for a wide range of parameters \cite{Silaev2} (see also remark \cite{remark5}).}

 \section{Coherence lengths in TCGL model and reduction to the single component GL theory in the
$T\rightarrow T_c$ limit.}
\subsection{The reduction argument in Ref. \onlinecite{kogan}}
{ As we discussed above in two band superconductors
 the broken symmetry is only $U(1)$ \cite{reservation}
 thus, by symmetry, {\it in the limit} $T\rightarrow T_c$, TCGL
expansion should be reduced to the
conventional text-book single-component GL theory. However the
reduction derivation presented in \cite{kogan} is principally
incorrect.} The crux of the argument presented in Ref.
\onlinecite{kogan} is that, the TCGL field equations [Eqs. (3,4)
in Ref.\onlinecite{kogan}]
\begin{eqnarray}\label{Eq:StartingEqs}
 a_1\Delta_1+b_1\Delta_1|\Delta_1|^2-\gamma\Delta_2-K_1\Pi^2\Delta_1&=&0\\\nonumber
 a_2\Delta_2+b_2\Delta_2|\Delta_2|^2-\gamma\Delta_1-K_2\Pi^2\Delta_2&=&0\\\nonumber
 \nabla^2 A-\nabla(\nabla\cdot A)&&\nonumber\\+\frac{16\pi^2}{\phi_0}\sum_{\nu=1,2}K_\nu(\Delta_\nu^*\Pi\Delta_\nu
 -\Delta_\nu(\Pi\Delta_\nu)^*)&=&0
 \end{eqnarray}
are {\it generically}, i.e. irrespectively of intercomponent
coupling strength $\gamma$ are well-approximated near $T_c$ by the
simpler system describing condensates with equal coherence length
{\it coupled only by a vector potential}  [Eqs. (7) and (8) in
Ref.\onlinecite{kogan}]
\begin{eqnarray}\label{Eq:ReducedEquations}
\alpha\tau\Delta_1+\beta_1\Delta_1|\Delta_1|^2-K\Pi^2\Delta_1&=&0\\
\alpha\tau\Delta_2+\beta_2\Delta_2|\Delta_2|^2-K\Pi^2\Delta_2&=&0\nonumber\\
\nabla^2 A-\nabla(\nabla\cdot A)&&\nonumber\\
+\frac{16\pi^2}{\phi_0}\sum_{\nu=1,2}K_\nu(\Delta_\nu^*\Pi\Delta_\nu
-\Delta_\nu(\Pi\Delta_\nu)^*)&=&0,
\end{eqnarray}
where $\Pi=\nabla - ie {\bf A}$ and the new parameters $\alpha, K,
\beta_\nu$ are related to the coefficients in starting
Eqs.(\ref{Eq:StartingEqs}) and GL parameter $\tau$.

Below we present a generic argument that { this result and
therefore the reduction procedure} are incorrect at any
temperatures. First we comment that the obtained reduced system of
Eqs.(\ref{Eq:ReducedEquations}) contradicts the basic principles of
Landau theory. That is, the initial set of equations corresponds
to the system with broken $U(1)$ symmetry. { Equations from the
second set} (\ref{Eq:ReducedEquations}) are coupled only through
${\bf A}$ and thus corresponds to independently conserved
condensates. Thus Eqs.(\ref{Eq:ReducedEquations}) are the field
equations corresponding to a free energy functional with
{spontaneously} broken $U(1)\times U(1)$ symmetry. Note that the
interband Josephson coupling in the initial set of equations
breaks the symmetry of the system down to $U(1)$ symmetry, but no
phase locking terms are present in the reduced system of
equations. Therefore the reduced theory fails to account to this
effect and is wrong already on symmetry grounds. 
Furthermore Landau theory for $U(1)$ systems dictates that
there is only one diverging coherence length {\it in the limit} $T
\to T_c$ associated with a single complex field (but not two
degenerate coherence lengths associated with two fields coupled
by vector potential only).

\subsection{Coherence lengths in two-band superconductors}

 {
The work \onlinecite{kogan} (and the recent follow up quoted therein \onlinecite{shanenko}) claim that the  gap fields $\Delta_{1,2}$ have two
independently diverging in the limit $T \to T_c$    coherence
lengths.  They claim that coherence lengths are directly attributed $\Delta_{1,2}$ and that they
 become degenerate in some domain, which  these authors call ``GL domain" at small $\tau$, where
the system is claimed to be described by two equations coupled
only by vector potential.}

{ Such incorrect conclusion regarding the evolution of the length
scales in the $T \to T_c$ limit is based on misunderstanding of
how coherence lengths are defined in two-band systems.} The
erroneous claim that two coherence lengths are attributed directly
to $\Delta_{1,2}$ and  that they become identical near $T_c$)
originates in the attempt \cite{kogan1} to assess {coherence
lengths though a comparison of the gap function profiles in the 1D
boundary problem. From the observation that the overall profiles
become identical in the $T \to T_c$ limit the authors of
Refs.\onlinecite{kogan1,kogan} concluded that the two gap
functions  $\Delta_{1,2}$ are characterized by the  similar coherence lengths.
Such approach is technically incorrect because one cannot extract
information of coherence length from the naive inspection of an
overall density profiles in a nonlinear theory. Instead the
correct analysis of coherence lengths in two-band superconductor
requires an accurate consideration of the asymptotic solutions of
linearized field equations for the gap
functions\cite{GL2mass2,Silaev}. As shown
\cite{GL2mass2,Silaev,Silaev2} in the wide range of parameters for
finite interband Josephson coupling there exist two asymptotical
normal modes with different coherence lengths (or inverse masses
of the normal modes). The two distinct coherence lengths appear as
a result of hybridization of the superconducting gap fields, and
{\it cannot} be directly attributed to the $\Delta_{1,2}$ fields
at any finite Josephson coupling and at any temperature. Instead
normal modes are associated with linear combinations of
$\Delta_{1,2}$ and thus coherence lengths are hybridized
\cite{GL2mass,GL2mass2,Silaev,Silaev2}. { Moreover one of the
two coherence lengths does not diverge in the limit $\tau \to 0$.
In fact the disparity between coherence lengths grows rather than shrinks in that limit \cite{Silaev,Silaev2}. 
This in particular means that type-1.5 regime in two-band superconductors cannot
 have anything in common  with a two-component counterpart of Bogomolnyi regime
 of single-component superconductors with $\kappa \approx 1/\sqrt{2}$ (see e.g. review \cite{newreview}).

The overall gap function profiles are determined by nonlinearities
and thus not only by masses of the normal modes but also by their
amplitudes \cite{Silaev2}. Therefore it is not possible to extract
the information about the coherence lengths just analyzing the
overall profile of the gap functions, like was attempted in Ref  \cite{kogan1}. The correct
reduction of TCGL model to single-component GL theory takes place
because in the limit $T \to T_c$ the mode with a non-diverging
coherence length looses its {\it amplitude} \cite{Silaev2}, but
not because two coherence lengths gradually become degenerate. }

\section{Misconceptions}

{\bf i}
For unclear reasons, the Ref. \onlinecite{kogan} criticises previous
works on type-1.5 superconductivity for {\sl
``assumption of two different penetration lengths
$\lambda_{1,2}$"}. We are not aware of {any} papers on two-band
superconductivity where such assumptions were made.
 As far as we know the notations $\lambda_{1,2}$ were used in literature on type-1.5
superconductivity only as characteristic constants, parameterizing
GL free energy
 while the physical magnetic field penetration length was always determined self-consistently.

{\bf ii} { In contrast to what was attributed to us in
Ref.\onlinecite{kogan1} no assumptions of having zero interband
coupling but equal $T_{c}$ for all components in two-band systems
were made \cite{bs1,GL2mass,GL2mass2,Silaev}.}

{\bf iii} { The work Ref.\onlinecite{kogan} claims that it is not
possible to obtain} fractional vortices in multicomponent
superconductors discussed in the Refs.\cite{frac}. Obviously {\it
in the limit} $T \to T_c$ in two-band systems fractional vortex
excitations are suppressed. However fractional vortices can exist
in $U(1) \times U(1)$ systems and in two-band $U(1)$ theories at
finite-$\tau$.  The authors of Ref.\onlinecite{kogan} missed that
the papers in Ref.\onlinecite{frac} deal with fractional vortex
solutions not in the $T \to T_c$ limit and not even in the GL
model but exclusively in the  London theory. In fact in a GL model
the fractional vortex solutions are quite different (see
corresponding discussion for $U(1)\times U(1)$ systems in
Ref.\onlinecite{juha}).
 Moreover Refs. \onlinecite{frac} primarily
focuses on the $U(1)\times U(1)$ systems. Thus the results in
Ref.\onlinecite{frac} are entirely unrelated to the arguments on
$T\to T_c$ limit in two-band systems. However we mention that
occurrence of fractional vortices in  two-component GL models with
intercomponent Josephson coupling in mesoscopic samples was
investigated by other groups \cite{fracmeso}.

{\bf iiii} {The work \cite{kogan} also contains mutually exclusive
claims. On one hand from the incorrect derivation of
Eqs.(\ref{Eq:ReducedEquations}) it would follow that in the limit
$T\rightarrow T_c$ the $U(1)$ TCGL theory is reduced to the
$U(1)\times U(1)$ theory when the gap functions are coupled only
by the vector potential (and not by  phase-locking terms). On the other
hand Ref. \cite{kogan} claims that in two-band superconductors the GL theory can
only describe the gap functions having the same phase. If the {
former} of these claims contradicts the basic principles of Landau theory
 (see the discussion above), the { latter} statement
also yields unreasonable conclusions negating for example the
existence of the phase difference excitations \cite{leggett}.}  At finite-$\tau$
when two-band GL theory is well justified, the appearance of the gradients
of the phase difference between component at finite $\tau$ is in
fact a quite generic effect   \cite{semimeissner} because the mass of the
phase-difference mode does not diverge in the $T \to T_c$ limit
\cite{Silaev2}.

\section{Conclusions}

We discussed the errors in the treatment of $T \to T_c$ limit { in
two-component superconductors} in Ref. \cite{kogan}, which led to the
incorrect (at any temperatures) system of field equations
(\ref{Eq:ReducedEquations}) { for the gap fields}  and incorrect
conclusions on the behavior of coherence lengths. We also pointed
out  that contrary to the claims in Ref.\onlinecite{kogan} TCGL
 expansion  is justified and can be used to describe systems
with disparity in coherence lengths as was demonstrated on formal grounds in \cite{Silaev2}
and does allow type-1.5 regimes.

{\it Note added}: below we also attach a separate rejoinder on reply by Kogan and Schmalian.

\section{Acknowledgements}
We thank Martin Speight, Alex Gurevich, Johan Carlstrom, Julien Garaud for discussions. This work was supported by
US National Science Foundation CAREER Award No. DMR-0955902,
by Knut and Alice Wallenberg Foundation through the Royal Swedish Academy of Sciences, Swedish Research Council,
 ``Dynasty" Foundation, Presidential RSS Council
(Grant No.MK-4211.2011.2).

\clearpage

\onecolumngrid

\begin{center}
\large{\textbf{
Rejoinder on  reply by
V. G. Kogan and J. Schmalian 
}}
\end{center}
\vspace{-0.25cm}
\begin{center}
Egor Babaev$^{1,2}$ and Mihail Silaev$^{2,3}$
\end{center}
\vspace{-0.3cm}
\begin{center}
$^{1}$\textit{Physics Department, University of Massachusetts, Amherst,  Massachusetts 01003, USA}  \\
$^{2}$ \textit{Department of Theoretical Physics, The Royal Institute of Technology, 10691 Stockholm, Sweden} \\
$^{3}$ \textit{Institute for Physics of Microstructures RAS, 603950 Nizhny Novgorod, Russia.}
\end{center}

\twocolumngrid

In their reply \cite{Rej:reply} to our recent comment \cite{Rej:comm},
Kogan and Schmalian did not  refute or, indeed,   discuss
 the main points of criticism in the comment. Unfortunately they instead advance
new  incorrect claims  regarding two-band and
type-1.5 superconductivity. The main purpose of this rejoinder is to
discuss these new incorrect claims.

{\bf (i)} First we would like to emphasize that 
in our comment we observed that the attempted GL construction 
in \cite{Rej:kogan} is incorrect {\em at all temperatures} and contradicts the basic
(symmetry-based) principles of Landau theory.
Thus it is not clear to us why in their reply the authors Ref.\onlinecite{Rej:reply} 
would assert that we ``share" the opinions espoused in
\cite{Rej:kogan}. Perhaps we did not make this point clear enough.
The standard understanding in the field
has always been indeed that a GL expansion to leading 
order in $\tau$  gives a single GL equation.
This elementary mean-field theory 
fact \cite{Rej:pathria} follows from simple symmetry considerations.
However, since the  mid-1960s it has also been known that, in principle, one can obtain
two-component GL equations from a multi-parameter expansion
in two-band systems \cite{Rej:Tilley} (see also remark \cite{Rej:rrr}.)
As we pointed out in our comment \cite{Rej:comm}, the picture
presented by Kogan and Schmalian in \cite{Rej:kogan} 
strongly disagrees with with the basic principles of GL theory:
 instead of obtaining
a standard single-component GL equation the authors of Ref.
\onlinecite{Rej:kogan} obtained a {\it system} of GL equations in the
leading order in $\tau=(1-T/T_c)$. This system of GL equations
corresponds to $U(1)\times U(1)$ broken symmetry which is
principally incorrect for two-band superconductors with  $U(1)$
broken symmetry. 

The correct derivation of a single GL equation in the \underline{ limit}
$\tau \to 0$, and a different system of two-component GL equations
for small but nonzero $\tau$, are discussed in Ref. \cite{Rej:Silaev2}. The authors
of \cite{Rej:reply} did not refute the criticism in our comment that,
in two-band superconductors, the two-component GL equation with different
coherence lengths is
rigorously obtained by a multiparameter expansion, and is not an
expansion in a single parameter $\tau$.
\begin{figure}[ht!]
\centerline{\includegraphics[width=1.0\linewidth]{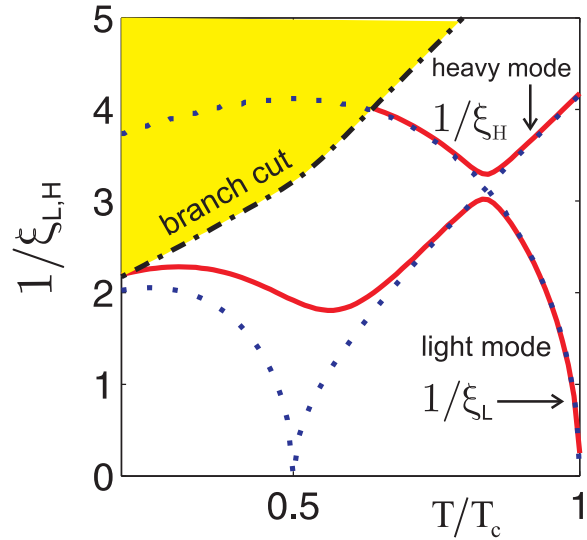}}
\caption{\label{Fig:modes} Comparison of the inverse coherence
lengths $\xi^{-1}_{L,H} (T)$ of a $U(1)\times U(1)$ model (dotted blue
lines) and a two-band $U(1)$ model with interband pairing (red
lines). The $U(1)\times U(1)$ model has $T_{c1}=1$ and  $T_{c2}=0.5$. In the
$U(1)\times U(1)$ case, one of the coherence lengths diverges at
$T_{c2}=0.5$. Above this temperature, for this component  we plot
the coherence length associated with superconducting fluctuations
in the normal state (around zero vacuum expectation value of
superconducting condensate for that component). It indeed
diminishes with increasing temperature. When a very small
interband coupling is added in this example, for obvious reasons,
it does not affect dramatically the characteristic length scales. It
leads only to a small hybridization of coherence lengths in this
case \cite{Rej:GL2mass,Rej:GL2mass2,Rej:Silaev,Rej:Silaev2}.
 However (i) the interband Josephson coupling removes a phase transition and a divergence of coherence length
at $T_{c2}$ and (ii) it gives both components nonzero expectation value above $T_{c2}$
without dramatically  affecting the characteristic length scales. Thus
there is growing disparity of coherence length near $T_c$ in such two-band superconductors.
Note  that these coherence lengths are associated with linear combinations of the gap fields. If one plots cores in the individual
gap fields or GL theory fields $\psi_{1,2}$, then,
because of nonlinear effects, the {\it overall density profile of vortex cores in $\psi_{1,2}$ will  become increasingly
similar near $T_c$  despite the divergence in the disparity of coherence lengths \cite{Rej:Silaev,Rej:Silaev2}}.
This is  because near $T_c$ there is less and less density associated with the mode which has short coherence length.
This effect was mistaken for being a signature that coherence lengths become similar in  the erroneous analysis in \cite{Rej:kogan1}.}
\end{figure}

{\bf (ii)} The authors of reply \cite{Rej:reply} incorrectly 
speculate how coherence lengths behave near $T_c$ in two-band superconductors, using their spurious
form of the GL equations.
They claim
that the difference in coherence lengths in two-band superconductors
near $T_c$ cannot be larger than $\tau^2$. This erroneous
conclusion is merely a consequence of the incorrect $U(1)\times U(1)$ form of their
``leading order in $\tau$" GL functional.
It is  possible to verify that it 
is indeed wrong by a comparison to the known (at all temperatures) behavior of coherence lengths in a full microscopic
theory. 
The Figure \ref{Fig:modes} shows the actual
behavior of two coherence lengths $\xi_{L,H}$ calculated in a
microscopic two-band Eilenberger theory \cite{Rej:Silaev,Rej:Silaev2} with interband coupling
  properly taken into account. Clearly the disparity of
$\xi_{L}$ and $\xi_{H}$ near $T_c$ does not disappear but instead
diverges  because only one coherence length $\xi_{L}$ diverges in
the limit $T\rightarrow T_c$.  This behavior is also
perfectly captured by the two-component GL theory \cite{Rej:Silaev2}.

Thus the behavior of coherence lengths proposed in the
reply\cite{Rej:reply}, not only is wrong on symmetry grounds but also
contradicts the results of microscopic theory
\cite{Rej:Silaev,Rej:Silaev2} which does not rely on any form of GL
expansion.

{\bf (iii)} Based on the erroneous behavior of coherence lengths
the authors of reply \cite{Rej:reply} further advance an obviously
incorrect claim that one should have $\kappa_1$ and $\kappa_2$ close to $1/\sqrt{2}$ in order to have type-1.5 behavior near
$T_c$. 
(We do not use the notations
$\kappa_1$ and $\kappa_2$, which also were not defined in \cite{Rej:reply};  we assume that they meant
$\kappa_1=\lambda/\xi_1$ and $\kappa_2=\lambda/\xi_2$.)
The authors of  \cite{Rej:reply} then assert that 
it is related to ``the situation studied by Essmann's group",  and ``the vortex attraction in small fields
for $\kappa$ close to $1/\sqrt{2}$ was not a sufficient reason for
declaring a new type of superconductivity".  In fact
Essmann's group studied a {\it single}-component type-2 superconductor 
which is close to the Bogomolnyi limit $\kappa \approx 1/\sqrt{2}$.
{These statements are obviously wrong since

({\bf a}.) In two-band systems
  the disparity  between $\xi_1$ and $\xi_2$ {\it diverges} near
$T_c$ in GL theory. Thus the type-1.5 regime has nothing to do with any kind of two-component
counterpart of the Bogomolnyi  ($\kappa \approx 1/\sqrt{2}$) regime.}

({\bf b}.)
As emphasized in our works, in single-component type-2
systems indeed
there is a well known effect of intervortex attraction due to
microscopic corrections in single-component theory with $\kappa
\approx 1/\sqrt{2}$. In that regime, at the level of GL theory
vortices almost do not interact and small microscopic corrections
become important. 
The physics of microscopic corrections in
single-component system is not universal (i.e. it is not based on
fundamental length scales but is a consequence of some particular,
e.g. weakly coupled BCS theory). Indeed by itself it
does not constitute a new regime of superconductivity since Nb 
studied by Essmann et.al. is still characterized by  a single ratio of
two fundamental length scales $\xi$ and $\lambda$ which categorizes it as a type-2 superconductor.

({\bf c}.) 
 The authors
of Ref.\onlinecite{Rej:reply} miss the fact that attractive
intervortex interaction is { not} a defining property of
``type-1.5 regime". 
All the physics
of type-1.5 superconductivity is about multi-component
superconducting states with $\xi_1 < \sqrt{2}\lambda< \xi_2$,  
and thus about the coexistence of  competing type-1 and
type-2 behaviors of {\it two} or more components (see e.g. the review \cite{Rej:review}).

Finally we do not agree with the assertion in \cite{Rej:reply} that the proceeding contributions of Brandt and
Das quoted therein provide an accurate review 
of the single-component $\kappa
\approx 1/\sqrt{2}$ regime,
 and unfortunately we had to comment on this
elsewhere \cite{Rej:bsnew}.

{(\bf iv)} In the reply \cite{Rej:reply} the authors  claim {\sl `` in
the statement that in ``the GL domain" the phases of two order
parameters must be the same modulo $\pi$, we had to stress that
this is true for the minimum energy state (equilibrium).
Otherwise, a reader may deduce that we negate the existence of the
Leggett mode as a possible excitation."}

First we note that this  contradicts what was stated in the original paper by Kogan and Schmalian
 \cite{Rej:kogan}. Namely in Ref. \cite{Rej:kogan} the argument that ``there is only one phase" was 
 used to claim that there are no fractional vortex excitations. These vortices
are nothing but phase difference excitations.

Second, the above-quoted statement is  self-contradictory. On one hand both in the
original paper \cite{Rej:kogan} and in the reply \cite{Rej:reply} the
authors claim the phases of the gap functions are locked in what
they call the ``GL domain". But on the other hand this statement
directly contradicts the system of equations which they write (eqs. (7),(8) in \cite{Rej:kogan}).
That system indeed does not have any phase locking terms at all.
Allowing phase difference variations in the
model\cite{Rej:kogan} does not yield a Leggett mode in contrast to
what was claimed in \cite{Rej:reply}.
 Namely  the system of two equations for the gap functions in
\cite{Rej:kogan}  does not contain the
interband coupling terms. The condensates are independently
conserved and thus such a system does not support  the Leggett 
mode which appears due to interband tunneling of Cooper pairs in
two-band superconductors. Instead in the incorrect GL construction in ref. \cite{Rej:kogan} the
phase difference mode is an unphysical Goldstone mode, existence of which   is
forbidden by symmetry in two-band superconductors. Thus   the
authors claim physical effects which are in direct contradiction
with the equations they write (phase locking effect in the absence of
any phase locking terms or Leggett mode in the absence of any interband
tunneling).


\begin{thebibliography}{99}
\bibitem{Rej:rrr}
 This kind of expansion was
not rigorously justified at the time of publication of the first paper
by Kogan and Schmalian \cite{Rej:kogan},
but it was rigorously justified in \cite{Rej:Silaev2} before the appearance
of the reply \cite{Rej:reply}.
\end{thebibliography}
\end{document}